\documentclass[
aps,
prl,
reprint,
superscriptaddress,
showpacs,
longbibligraphy
]{revtex4-2}
\usepackage{xcolor}
\usepackage{graphicx}
\usepackage{amsmath, amssymb, amsfonts, bm, bbm}
\usepackage{natbib}
\usepackage{hyperref}
\usepackage{dcolumn}
\usepackage{xcolor}
\hypersetup{colorlinks=true,linkcolor=blue,urlcolor=blue,citecolor=blue}
\usepackage[normalem]{ulem}

\begin{document}


\title{Continuum theory for topological phase transitions in exciton systems}


\author{Xiaochan Cai}
\affiliation{School of Physical Science and Technology, ShanghaiTech University, Shanghai 201210, China}
\affiliation{Institut für Theoretische Physik und Astrophysik and Würzburg-Dresden Cluster of Excellence ct.qmat, Universität Würzburg, 97074 Würzburg, Germany}

\author{Armando Consiglio}
\affiliation{Institut für Theoretische Physik und Astrophysik and Würzburg-Dresden Cluster of Excellence ct.qmat, Universität Würzburg, 97074 Würzburg, Germany}
\affiliation{Istituto Officina dei Materiali, Consiglio Nazionale delle Ricerche, Trieste I-34149, Italy}

\author{Domenico Di Sante}
\affiliation{Department of Physics and Astronomy, University of Bologna, 40127 Bologna, Italy}

\author{Ronny Thomale}
\affiliation{Institut für Theoretische Physik und Astrophysik and Würzburg-Dresden Cluster of Excellence ct.qmat, Universität Würzburg, 97074 Würzburg, Germany}

\author{Werner Hanke}
\affiliation{Institut für Theoretische Physik und Astrophysik and Würzburg-Dresden Cluster of Excellence ct.qmat, Universität Würzburg, 97074 Würzburg, Germany}


\date{\today}

\begin{abstract}
An effective continuum theory is constructed for the topological phase transition of excitons in quasi-two-dimensional systems. These topological excitons crucially determine the optoelectronic properties, because of their larger binding energies in 2D as well as their topologically enhanced exciton transport.
The core idea of this letter is, that the essential physics determining the topological invariants across the phase transition is localized near $N$-fold band-crossing points (BCPs) in the interaction-induced exciton band structure.  The construction of the continuum theory around these BCPs needs only the information of exciton states that build up these BCPs at both $\mathbf{Q}=0$ and finite $\mathbf{Q}$ points, and not the numerically challenging solution of the Bethe-Salpeter equation over the full exciton Brillouin zone. This theory applies to systems with and without spin conservation.
Our theory is illustrated in two specific examples: the transition metal dichalcogenide twisted bilayer systems and the Bernevig-Hughes-Zhang (BHZ) model.
These results offer a promising route toward studying complex systems, such as the room-temperature quantum spin Hall system Bismuthene (Bi/SiC) and other twisted bilayer systems.
\end{abstract}

\maketitle


\textit{Introduction}.---Excitons are bound states of electrons excited into conduction bands and holes left behind in valence bands, which arise from an intricate interplay between attractive and repulsive forces, encoded in the Bethe-Salpeter equation (BSE), and are known to dominate the optoelectronic properties of a wide range of materials \cite{
Martin_Reining_Ceperley_2016,
10.1103/physrevlett.33.582,
PhysRevB.21.4656,
10.1080/00018737800101384,
10.1103/revmodphys.74.601,
10.1103/revmodphys.90.021001}.
Of particular interest are here topological excitons in quasi-two-dimensional (2D) systems, because of both the reduced screening (i.e., larger binding energies) and the enhanced transport characteristics \cite{10.48550/arxiv.2406.11951,2410.00967} inherent in their topology.
Our own interest partly stems also from our recent observation \cite{10.1038/s41467-022-33822-8} of room temperature excitons in an atomically thin topological insulator, Bismuthene (Bi/SiC), which we had in earlier work \cite{10.1126/science.aai8142,10.1103/physrevb.98.165146,10.1103/physrevb.98.161407} established as a large gap ($\sim0.8eV$) 2D quantum spin Hall (QSH) insulator \cite{10.1021/acs.jpcc.2c05809}.
A standard procedure to identify exciton topology involves computing topological invariants via the Berry-curvature integral or Wannier center evolution, methods traced back to electron topology \cite{10.1103/physrevlett.49.405,10.1016/0003-4916(85)90148-4,10.1143/jpsj.74.1674,10.1143/jpsj.76.053702,10.1103/physrevb.74.195312,10.1103/physrevb.83.235401,10.1103/physrevb.84.075119,10.1080/00018732.2015.1068524,10.1103/physrevb.89.155114}. These approaches have been applied to study topological excitons in Moiré heterostructures \cite{10.1103/physrevlett.126.137601,10.1103/physrevlett.118.147401,10.1103/physrevlett.133.136403,PhysRevResearch.7.023047,10.48550/arxiv.2504.10189} and theoretical models \cite{10.1103/physrevb.102.035146,10.48550/arxiv.2406.11951,PhysRevLett.133.176601}.
However, the need to solve the BSE over the full exciton Brillouin zone (XBZ) can be numerically rather demanding.

The central theme of our letter is that, when an exciton system approaches a topological phase transition, marked by local exciton-band-gap closure, such numerical complications can be strongly reduced by using a continuum theory description of the gap-closing points: the changes in topological invariants across the transition depend exclusively on the topological charges \cite{10.1093/acprof:oso/9780199564842.001.0001} of these points.
This approach is well studied in electronic systems \cite{10.1103/physrevb.76.205304,10.1103/physrevb.77.235125,10.1073/pnas.1308853110,10.1103/physrevb.85.195320,10.1073/pnas.1308853110,10.1103/physrevlett.102.096801,10.1103/physrevlett.100.036804,10.1126/science.1133734}, exemplified by the Bernevig-Hughes-Zhang (BHZ) model proposed to explain the topological features observed in the first experimental realization of HgTe quantum spin Hall (QSH) insulators \cite{10.1126/science.1133734,10.1126/science.1148047}, but remains relatively unexplored for excitons.
While several studies have developed continuum theories for exciton systems \cite{10.1038/ncomms4876, 10.1103/physrevb.89.205303, 10.1103/physrevb.91.075310,10.1103/physrevlett.115.176801,10.1103/physrevlett.118.147401,10.1038/srep42390,10.1103/physrevlett.133.136403,10.48550/arxiv.2504.10189}, most of them concentrate on the vicinity of the $\mathbf{Q}=0$ point, particularly in transition metal dichalcogenide (TMD) monolayers and twisted multilayers, systems that typically exhibit spin conservation. To date, continuum-theory studies in systems without spin conservation (e.g., Bi/SiC) have yet to be reported.
A recent study by Das Sarma's group \cite{10.1103/physrevlett.133.136403} extended previous works beyond the vicinity of $\mathbf{Q}=0$ by exploiting the strong XBZ folding induced by the Moiré potential, which effectively preserves the exciton character across the Moiré XBZ. However, when the exciton character varies significantly, a localized continuum description around discrete $\mathbf{Q}$ points becomes necessary.

In this letter, we present a general continuum theory for exciton bands near $N$-fold band-crossing points (BCPs), which applies to BCPs at both $\mathbf{Q}=0$ and finite $\mathbf{Q}$ points, and to systems with and without spin-conservation.
We focus here on systems that have time-reversal symmetry (TRS) at the transition point. Breaking the TRS in a topological trivial exciton system by applying, e.g., external magnetic fields, can drive it into a nontrivial phase, whose exciton Chern number can indeed be directly determined by the topological charges of the BCPs. This allows then for a consistency check between the numerical Berry-curvature integrals and the continuum approach [see Eq.~(\ref{eq-eg-TMD}) and Eq.~(\ref{eq-BHZ-exeffH}) below].

We apply our theory to two representative examples.
The first is the TMD twisted bilayer proposed by MacDonald's group \cite{10.1103/physrevlett.118.147401}. The lowest two exciton bands touch only at the $\mathbf{Q}=0$ point, representing a single-BCP case, and become detached via an external Zeeman term, making the system an ideal platform to test our theory.
Our second example concerns the BHZ model. In the QSH regime, the two lowest $S_z=0$ exciton bands form two BCPs at $\mathbf{Q}=0$ and a finite $\mathbf{Q}$ point.
Both BCPs can be gapped out by including both spin and orbital Zeeman couplings. Our continuum analysis shows that under the Zeeman terms, the lowest $S_z = 0$ exciton band acquires a nonzero exciton Chern number, consistent with our full numerical BSE calculations. Furthermore, we find this nontrivial topology specifically stems from the exchange electron-hole (e-h) interaction.

These two examples represent a first step to verify our theory, which, however, extends to a broader class of systems, such as the twisted bilayer structures of TMDs and graphene \cite{10.1103/physrevlett.126.137601}, as well as Chern insulators \cite{10.1103/physrevb.96.161101}. Additionally, this approach offers a promising route to investigate exciton topology in more complex systems, like Bi/SiC, where intricate band structures pose significant challenges for conventional methods.

\textit{BSE and exciton Chern number}.---Excitons can generically be expressed as linear superpositions of electron-hole (e-h) pairs
\begin{equation}
\begin{aligned}
   |\Psi_{\mathbf{Q}}\rangle =\sum_{\mathbf{k}vc} \phi_{\mathbf{Q}vc}(\mathbf{k}) |vc\mathbf{k}\mathbf{Q}\rangle,
\end{aligned}
\end{equation}
where the sum runs over all valence ($v$) and conduction ($c$) bands, and $\mathbf{Q}$ and $\mathbf{k}$ are the center-of-mass and relative momenta, respectively. The e-h pair $|{v}{c}\mathbf{k}\mathbf{Q}\rangle=|\psi_{c,\mathbf{k}+\mathbf{Q}}\rangle|\psi_{v,\mathbf{k}}\rangle^*$ consists of an electron in $|\psi_{c,\mathbf{k}+\mathbf{Q}}\rangle$ and a hole in $|\psi_{v,\mathbf{k}}\rangle$, with the amplitude given by the envelope function $\phi_{\mathbf{Q}vc}(\mathbf{k})$.
For a system described by the Hamiltonian $\mathcal{H}=\mathcal{H}_0+\mathcal{H}_V$, where $\mathcal{H}_0$ and $\mathcal{H}_V$ are the non-interacting and interaction parts, the exciton band structure is obtained by solving the eigenvalue problem for the Hamiltonian matrix $
\langle{v}{c}\mathbf{k}\mathbf{Q}|\mathcal{H}|{v}^{\prime}{c}^{\prime}\mathbf{k}^{\prime}\mathbf{Q}\rangle$, or equivalently, by solving the BSE \cite{10.1103/revmodphys.74.601}
\begin{equation}\label{eq-BSE}
\begin{aligned}
    &\left(\varepsilon_{c,\mathbf{k}+\mathbf{Q}}-\varepsilon_{v,\mathbf{k}}\right)
    \phi_{\mathbf{Q}vc}(\mathbf{k})
    \\
    &\quad+
    \sum_{\mathbf{k}^{\prime}v^{\prime}c^{\prime}}
    \langle{v}{c}\mathbf{k}\mathbf{Q}|\mathcal{K}|{v}^{\prime}{c}^{\prime}\mathbf{k}^{\prime}\mathbf{Q}\rangle
    \phi_{\mathbf{Q}v^{\prime}c^{\prime}}(\mathbf{k}^{\prime})
    =E_{\mathbf{Q}}\phi_{\mathbf{Q}v^{\prime}c^{\prime}}(\mathbf{k}^{\prime}).
\end{aligned}
\end{equation}
Here $E_{\mathbf{Q}}$ is the exciton eigen-energy, $\varepsilon_{c,\mathbf{k}}$ ($\varepsilon_{v,\mathbf{k}}$) is the quasiparticle energy of $|\psi_{c,\mathbf{k}}\rangle$ ($|\psi_{v,\mathbf{k}}\rangle$), and the interaction kernel
$\langle{v}{c}\mathbf{k}\mathbf{Q}|\mathcal{K}|{v}^{\prime}{c}^{\prime}\mathbf{k}^{\prime}\mathbf{Q}\rangle=\langle{v}{c}\mathbf{k}\mathbf{Q}|\mathcal{K}^{x}+\mathcal{K}^{d}|{v}^{\prime}{c}^{\prime}\mathbf{k}^{\prime}\mathbf{Q}\rangle$ can be decomposed into the exchange term $\mathcal{K}^{x}$ and the direct term $\mathcal{K}^{d}$.
Eq.~(\ref{eq-BSE}) is fundamental for describing excitons and their associated physical phenomena, including their topological properties.
For a set of exciton bands isolated from the others, the topology of this set is characterzied by the exciton Chern number, which is defined similarly to that in electron systems \cite{Bernevig-2013}
\begin{equation}\label{eq-exciton-Chern}
    \mathcal{C}_{\text{exc}} = \frac{1}{2\pi}\sum_{n}\int_{\text{XBZ}} d^2\mathbf{Q}[\mathbf{\Omega}^{n}_{\mathbf{Q}}]_z.
\end{equation}
Here $n$ runs over all bands in the set, $\mathbf{\Omega}^{n}_{\mathbf{Q}}=\nabla_{\mathbf{Q}}\times\mathcal{A}^{n}_{\mathbf{Q}}$ and $\mathcal{A}^{n}_{\mathbf{Q}}=i\langle\Psi^{n}_{\mathbf{Q}}|\nabla_{\mathbf{Q}}|\Psi^{n}_{\mathbf{Q}}\rangle$ are the Berry curvature and the Berry connection of the exciton band $|\Psi^{n}_{\mathbf{Q}}\rangle$, respectively, both of which can be extracted from Eq.~(\ref{eq-BSE}).

\begin{figure}[b]
    \centering
    \includegraphics[width=0.5\textwidth]{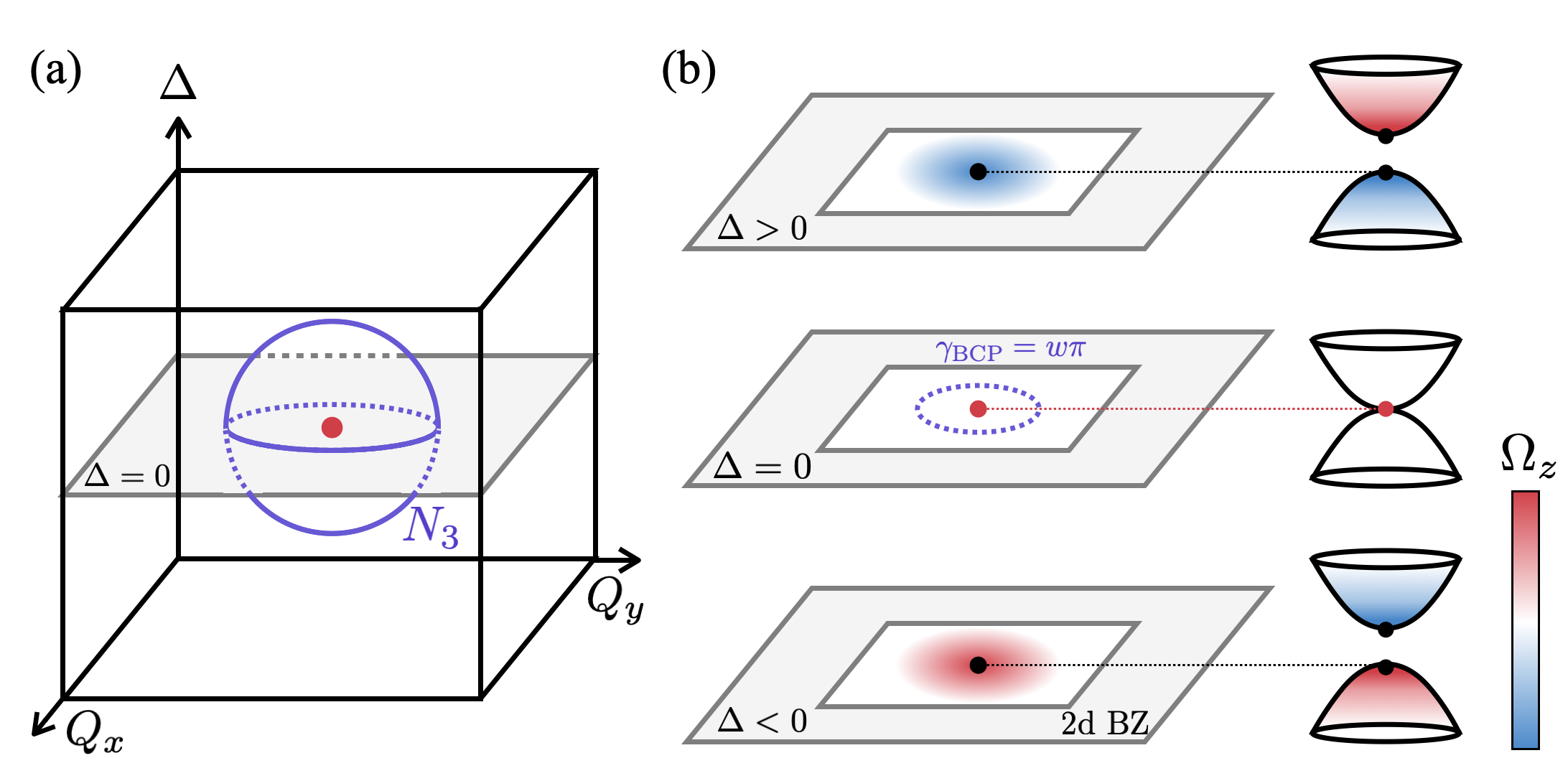}%
    \caption{
    (a) In the 3D parameter space $(Q_x,Q_y,\Delta)$, the topological charge $N_3$ of a BCP (marked by the red dot on the $\Delta=0$ plane) at the transition is obtained by integrating exciton Berry curvature $\mathbf{\Omega}(Q_x,Q_y,\Delta)$ over the closed surface (marked by purple sphere).
    (b) A non-zero $N_3$ manifests itself as the discontinuous change of Berry-curvature distribution around the BCP. When TRS is present, $N_3$ is related to the Berry phase through $N_3=\frac{\gamma_{\text{BCP}}}{\pi}=w$.
    }
    \label{fig-general-scheme}
\end{figure}

\textit{Topological phase transition and continuum theory for excitons.}---The continuum theory approach was first introduced in electron systems and thoroughly discussed by Bernevig, Hughes, and Zhang in their celebrated work on HgTe quantum wells \cite{10.1126/science.1133734}. The core idea, which applies equally to excitons, is founded on the principle that, during a topological phase transition, the essential physics responsible for the jump in the topological invariant is localized around the gap-closing points and can thus be faithfully captured within a continuum theory framework.
For excitons, these gap-closing points manifest as the $N$-fold BCPs formed by $N$ exciton bands.
When the phase transition is driven by tuning a parameter $\Delta$, with $\Delta=0$ set as the transition point, the difference in $\mathcal{C}_{\text{exc}}$ on both sides is given by
\begin{equation}\label{eq-Chern-diff-topological-charge}
    \mathcal{C}_{\text{exc}}(\Delta>0)-\mathcal{C}_{\text{exc}}(\Delta<0) = \sum_{i}N^{i}_3,
\end{equation}
where the sum goes over all BCPs and $N^{i}_3 =\frac{1}{2\pi} \oint_{S_i}d\mathbf{S}\cdot\mathbf{\Omega}(Q_x,Q_y,\Delta)$ is the topological charge of the $i$th BCP. Here the exciton Berry curvature $\mathbf{\Omega}(Q_x,Q_y,\Delta)$ is defined in the extended 3d parameter space $(Q_x,Q_y,\Delta)$, and $S_i$ is a closed surface surrounding the $i$th BCP [see Fig.~\ref{fig-general-scheme}(a)].
In this way, the problem of calculating $\mathcal{C}_{\text{exc}}$ naturally reduces to the task of constructing the effective continuum Hamiltonian \cite{topological-charge}.

\begin{figure}[tb]
    \includegraphics[width=1\linewidth]{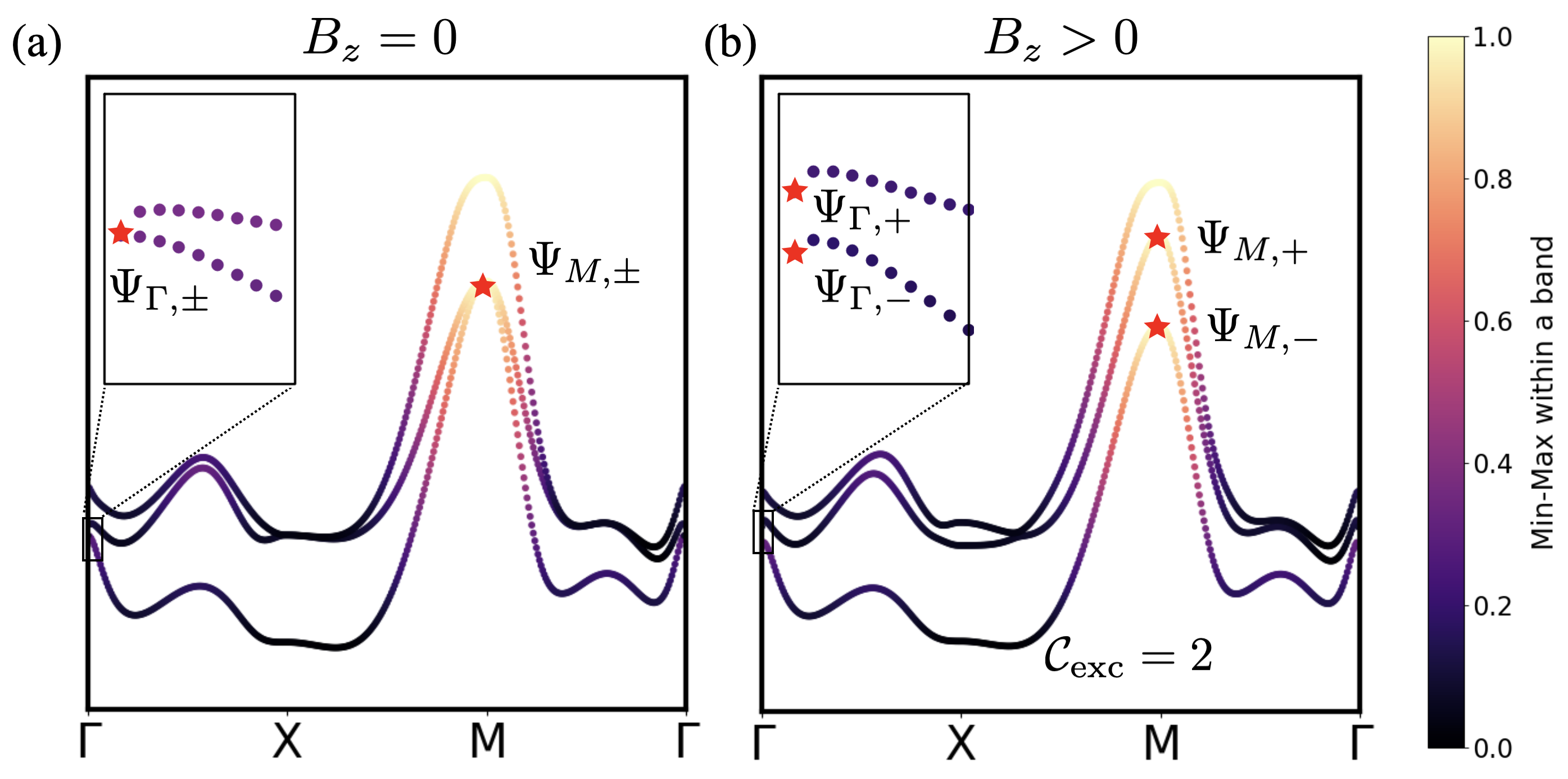}
    \caption{The $S_z=0$ exciton dispersions of the BHZ model (a) without and (b) with the Zeeman term $H_{B_z}$. Red stars mark the two BCPs lifted upon applying $H_{B_z}$. When $B_z>0$, the lowest exciton band becomes fully isolated and acquires the exciton Chern number $\mathcal{C}_{\text{exc}}=2$. See Sec.~III.A of the Supplementary Material \cite{SI} for numerical details.}
    \label{fig-BHZ-exciton-dispersion}
\end{figure}

Existing works of continuum theory for excitons have primarily focused on BCPs at $\mathbf{Q}=0$ in spin-conserved systems \cite{continuum-theory}.
Here we present a general continuum theory framework for an N-fold BCP (details are provided in Sec.~I of the Supplementary Material \cite{SI}), extending the scope to two less-explored yet physically relevant cases: (i) BCPs at finite-$\mathbf{Q}$ points \cite{10.1103/physrevb.96.161101}, and (ii) systems without spin-conservation.
Representative examples exist for both cases: the former is realized by $S_z=0$ excitons in the BHZ model, which will be revisited in our second example below [see Fig.~\ref{fig-BHZ-exciton-dispersion}(a)]; the latter is exemplified by Bi/SiC, where Rashba spin–orbit coupling (SOC) induces spin mixing \cite{10.1126/science.1133734,10.1126/science.1148047}.
Consider a BCP formed by $N$ exciton states $\{|\Psi^{\mu}_{\mathbf{Q}_0}\rangle\}$ at $\mathbf{Q}_0$ point.
The exciton continuum Hamiltonian is an $N\times{N}$ matrix, given as
\begin{equation}
    [H^{\text{eff}}_{\mathbf{Q}_0}(\mathbf{Q})]_{\mu\nu}=\langle\Psi^{\mu}_{\mathbf{Q}_0}(\mathbf{Q})|\mathcal{H}|\Psi^{\nu}_{\mathbf{Q}_0}(\mathbf{Q})\rangle,
\end{equation}
where $\mathbf{Q}$ is the momentum measured from $\mathbf{Q}_0$, $\mathcal{H}$ is the two-body Hamiltonian, $|\Psi^{\mu}_{\mathbf{Q}_0}(\mathbf{Q})\rangle=e^{i\mathbf{Q}\cdot\mathbf{R}}|\Psi^{\mu}_{\mathbf{Q}_0}\rangle$ and $\mathbf{R}$ is the center-of-mass coordinate.
The subsequent treatment of $|\Psi^{\mu}_{\mathbf{Q}_0}(\mathbf{Q})\rangle$ depends on whether the electron bands involved in $\{|\Psi^{\mu}_{\mathbf{Q}_0}\}\rangle$ are degenerate.
We notice that a previous study \cite{10.1103/physrevlett.115.176801} adopted an approach valid only for Frenkel excitons originating from non-degenerate electron bands.
Our treatment applies to both Wannier and Frenkel excitons.
Since two examples discussed below correspond to the non-degenerate case, we focus here on the resulting Hamiltonian matrix for this situation. A detailed discussion covering both degenerate and non-degenerate cases is presented in Sec.~I of \cite{SI}.
The continuum Hamiltonian can be expressed as the sum of three contributions: $H^{\text{eff}}_{\mathbf{Q}_0}(\mathbf{Q})=H^{sp}_{\mathbf{Q}_0}(\mathbf{Q})+K^{x}_{\mathbf{Q}_0}(\mathbf{Q})+K^{d}_{\mathbf{Q}_0}(\mathbf{Q})$, where $H^{sp}$, $K^{x}$ and $K^{d}$ are single-particle, exchange and direct terms, respectively.
In the \textit{ab initio} formalism (see Sec.~I of \cite{SI} for the tight-binding formalism), the exchange and direct terms read
\begin{equation}\label{eq-ex-di-term1}
\begin{aligned}
    [K^{x}_{\mathbf{Q}_0}(\mathbf{Q})]_{\mu\nu}
    &=\sum_{\mathbf{G}}
    V_{\mathbf{G}+\mathbf{Q}_0+\mathbf{Q}}
    D_{\mathbf{Q}_0\mu}(\mathbf{Q},\mathbf{G})
    D^{*}_{\mathbf{Q}_0\nu}(\mathbf{Q},\mathbf{G}),
    \\
    [K^{d}_{\mathbf{Q}_0}(\mathbf{Q})]_{\mu\nu}
    &=-
    \sum_{\mathbf{G}\mathbf{G}^{\prime}}
    M_{\mathbf{Q}_0\mu\nu}(\mathbf{Q},\mathbf{G},\mathbf{G}^{\prime}),
\end{aligned}
\end{equation}
where $\mathbf{Q}$ is the momentum measured from $\mathbf{Q}_0$, $\mathbf{G}$ is the reciprocal lattice vector, and
\begin{equation}\label{eq-ex-di-term2}
\begin{aligned}
    &D_{\mathbf{Q}_0\mu}(\mathbf{Q},\mathbf{G})
    =\sum_{\mathbf{k}vc}\phi^{\mu,*}_{\mathbf{Q}_0vc}(\mathbf{k})U^{c,\mathbf{k}+\mathbf{Q}_0+\mathbf{Q}/2}_{v,\mathbf{k}-\mathbf{Q}/2}(\mathbf{G})
    \\
    &M_{\mathbf{Q}_0\mu\nu}(\mathbf{Q},\mathbf{G},\mathbf{G}^{\prime})
    =
    \sum_{\mathbf{k}vc\mathbf{k}^{\prime}v^{\prime}c^{\prime}}
    W_{\mathbf{G}\mathbf{G}^{\prime}}(\mathbf{k}^{\prime}-\mathbf{k})
    \phi^{\mu,*}_{\mathbf{Q}_0vc}(\mathbf{k})
    \\
    &\quad\times
    \phi^{\nu}_{\mathbf{Q}_0v^{\prime}c^{\prime}}(\mathbf{k}^{\prime})
    {U}^{c,\mathbf{k}+\mathbf{Q}_0+\mathbf{Q}/2}_{c^{\prime},\mathbf{k}^{\prime}+\mathbf{Q}_0+\mathbf{Q}/2}(-\mathbf{G})
    U^{v^{\prime},\mathbf{k}^{\prime}-\mathbf{Q}/2}_{v,\mathbf{k}-\mathbf{Q}/2}(\mathbf{G}^{\prime}).
\end{aligned}
\end{equation}
In Eq.~(\ref{eq-ex-di-term1}) and Eq.~(\ref{eq-ex-di-term2}), $V$ ($W$) is the Fourier transformed bare (screened) Coulomb interaction,  $\phi^{\mu}_{\mathbf{Q}_0vc}(\mathbf{k})$ is the envelope function of $|\Psi^{\mu}_{\mathbf{Q}_0}\rangle$
and $U^{n,\mathbf{k}}_{n^{\prime},\mathbf{k}^{\prime}}(\mathbf{G})=\langle{u}_{n,\mathbf{k}}|e^{i\mathbf{G}\cdot\mathbf{r}}|u_{n^{\prime},\mathbf{k}^{\prime}}\rangle$, where $|u_{n,\mathbf{k}}\rangle=e^{-i\mathbf{k}\cdot\mathbf{r}}|\psi_{n,\mathbf{k}}\rangle$ ($n=c,v$) is the periodic part of the electron Bloch function.
We note that, for a generic $\mathbf{Q}_0$ point, $H^{\text{eff}}_{\mathbf{Q_0}}(\mathbf{Q})$ still needs to be solved numerically.
In contrast, when $\mathbf{Q}_0$ lies at a high-symmetry point, symmetry analysis alone can constrain the form of $H^{\text{eff}}_{\mathbf{Q_0}}(\mathbf{Q})$, allowing it to be solved analytically.
As will be shown below, the continuum Hamiltonians of the BHZ model are derived precisely in this way --- based on the constraints imposed by fourfold rotational symmetry and TRS, which is sufficient to capture the essential topological properties of excitons.

\textit{Nontrivial exciton topology induced by breaking TRS.}---
While significantly reducing the computational cost, the continuum theory approach comes with the limitation that, generally it can only determine the difference in $\mathcal{C}_{\text{exc}}$ across the transition [see Eq.~(\ref{eq-Chern-diff-topological-charge})] but not its absolute value \cite{chern-diff}.
This limitation can be circumvented by introducing a domain wall \cite{kink-state}, where the parameter $\Delta$ changes sign in real space. As such, a special type of edge state, called kink state, appears at the domain wall, whose number is directly determined by the topological charges of the BCPs.
Nevertheless, \textit{the overall topology of the system still remains undetermined}.
Here we adopt a different strategy --- explicitly breaking a specific symmetry to resolve the ambiguity in determining the exact value of $\mathcal{C}_{\text{exc}}$.
In particular, we focus on TRS, which is ubiquitous in non-magnetic materials and can be precisely broken, e.g., by applying an external magnetic field.
When BCPs are gapped by breaking TRS (i.e., $\Delta$ represents a TRS-breaking term), the two sides of the transition are related by TRS, enforcing $\mathcal{C}_{\text{exc}}(\Delta>0)=-\mathcal{C}_{\text{exc}}(\Delta<0)$. Combined with Eq.~(\ref{eq-Chern-diff-topological-charge}), one obtains
\begin{equation}\label{eq-Chern-topological-charge}
    \mathcal{C}_{\text{exc}}(\Delta) = \frac{1}{2}\sum_{i}N^{i}_3\cdot\operatorname{sgn}(\Delta),
\end{equation}
where $i$ goes over all BCPs and $N^{i}_3$ can be obtained from the effective continuum Hamiltonian around the $i$th BCP.
An important consequence of breaking TRS is that it lifts the topological constraint imposed by TRS: when TRS is present, $\mathcal{C}_{\text{exc}}$ must vanish. In contrast, breaking TRS can drive the system into a topologically nontrivial phase [see Eq.~(\ref{eq-Chern-topological-charge})]. Next, we will explicitly demonstrate this in the case of two-fold BCPs.

For a two-fold BCP located at $\mathbf{Q}_0$ point, the effective Hamiltonian in its vicinity takes the following form
\begin{equation}\label{eq-model-generic}
    H^{\text{eff}}_{\mathbf{Q}_0}(\mathbf{Q}) = \left(C+S_{\mathbf{Q}}+A_{\mathbf{Q}}\right)\cdot{I}_2+\left[
        \begin{array}{cc}
            \Delta & B_{\mathbf{Q}} \\
            B^*_{\mathbf{Q}} & -\Delta
        \end{array}
    \right],
\end{equation}
where $\mathbf{Q}$ is the momentum measured from $\mathbf{Q}_0$, $I_2$ is the $2\times2$ identity matrix and the coefficients have different origins \cite{2fold-coeffs}.
We shall set $C=S_{\mathbf{Q}}=A_{\mathbf{Q}}=0$, as they do not affect the topological properties of the system.
The parameter $\Delta$ acts as a Haldane mass term that breaks TRS, which can be controlled by a perpendicular external magnetic field, and opens a gap at the BCP.
$B_{\mathbf{Q}}$ generically depends on both the magnitude $Q$ and the orientation angle $\phi_{\mathbf{Q}}$ of $\mathbf{Q}$.
For simplicity, we assume an isotropic exciton dispersion around $\mathbf{Q}=0$ in the main text (see Sec.~II of \cite{SI} for the anisotropic case), resulting in
\begin{equation}\label{eq-BQ}
    B_{\mathbf{Q}}=b(Q)e^{-iw\phi_{\mathbf{Q}}},
\end{equation}
where $b(Q)$ is a real scalar function of $Q$ and $w$ is an integer-valued winding number associated with the phase of $B_{\mathbf{Q}}$. We note that Eq.~(\ref{eq-BQ}) applies to a wide range of material systems \cite{isotropic-assumption}.
In this case, the BCP carries a Berry phase of $\gamma_{\text{BCP}}=w\pi$ and a topological charge
\begin{equation}\label{eq-N3-w}
    N_3 = \frac{\gamma_{\text{BCP}}}{\pi}=w.
\end{equation}
Moreover, the local topological structure in its vinicity is described by the Berry curvature $[\Omega_{\mathbf{Q}}]_z = \frac{1}{4}\Delta{w}{Q}^{-1}\frac{d|b(Q)|^2}{dQ}\epsilon^{-\frac{3}{2}}_{\mathbf{Q}}$, where $\epsilon_{\mathbf{Q}}=\sqrt{|B_{\mathbf{Q}}|^2+\Delta^2}$.
As shown in Fig.~\ref{fig-general-scheme}(b), when $N_3$ ($w$) is nonzero, the local topological structure near $\mathbf{Q}=0$ undergoes a discontinuous change during the transition.
Combined with Eq.~(\ref{eq-Chern-topological-charge}), the exciton Chern number takes the form $\mathcal{C}_{\text{exc}}(\Delta)=\frac{1}{2}\sum_iw_i\cdot\operatorname{sgn}(\Delta)$.

Based on the above discussion, there are two key ingredients in exciton topology: the coefficient $B_{\mathbf{Q}}$ with \textit{a nonzero winding number} $w$ and the \textit{tunable mass term} $\Delta$, irrespective of their physical origins \cite{two-fold-case}.
Notably, in certain systems, the nonzero winding number can be reflected in optical responses. One classic example is the TMD monolayer, where a previous study demonstrated the optical generation of excitonic valley coherence \cite{10.1038/nnano.2013.151}, which can actually be interpreted as a manifestation of the two-fold BCP at $\mathbf{Q}=0$ carrying a winding number $w=2$. More importantly, in our first example, i.e., the TMD twisted bilayer, the introduction of the Moiré potential leads to a single BCP at $\mathbf{Q}=0$ in the low-energy exciton bands. When an external magnetic field opens a gap at this BCP, the topology of the lowest exciton band is uniquely determined by this $\mathbf{Q}=0$ BCP carrying $w=2$.
Finally, we note that the above conclusions [Eqs.~(\ref{eq-model-generic})–(\ref{eq-N3-w})] also hold for systems without TRS \cite{two-fold-case}.

\textit{Examples.}---We now apply our theory to two representative systems.
The first example is the TMD twisted bilayer studied by MacDonald's group \cite{10.1103/physrevlett.118.147401}.
The lowest exciton Moiré band crosses only with the second-lowest band at the $\gamma$ point of the Moiré XBZ, corresponding to a single-BCP case.
The effective exciton Hamiltonian around $\gamma$ carries a winding number $w=2$
\begin{equation}\label{eq-eg-TMD}
\begin{aligned}
    H^{\text{eff}}_{\gamma}(\mathbf{Q})
    = [\hbar\Omega_0&+\frac{\hbar^2Q^2}{2M}+JQ+\Delta(\mathbf{r})]\cdot{I}_2
    \\
    &\quad\quad+\left[
    \begin{array}{cc}
        h_z & JQe^{-i2\phi_{\mathbf{Q}}} \\
        JQe^{i2\phi_{\mathbf{Q}}} & -h_z
    \end{array}
    \right].
\end{aligned}
\end{equation}
Here $\hbar\Omega_0$ is the exciton energy at $\mathbf{Q}=0$, $\hbar^2Q^2/2M$ is the center-of-mass kinetic energy, $J$ is the coefficient for the exchange interaction contribution, $\Delta(\mathbf{r})$ denotes the Moiré superlattice potential and $h_z$ accounts for the Zeeman term steming from the magnetic field.
By setting $h_z>0$, a gap opens at this BCP and the lowest exciton band acquires a Chern number $\mathcal{C}_{\text{exc}}=1$.
Although Ref.~\cite{10.1103/physrevlett.118.147401} identifies the nonzero winding number as a potential mechanism for realizing topological exciton bands, the exciton Chern number is determined via the direct integration of Berry curvature. The off-diagonal matrix elements in Eq.~(\ref{eq-eg-TMD}) indicate that the BCP at the $\gamma$ point carries $N_3=2$ [see Eq.~(\ref{eq-N3-w})].
Consequently, upon applying a magnetic field $h_z$, the lowest Moiré exciton band acquires $\mathcal{C}_{\text{exc}}(h_z)=\operatorname{sgn}(h_z)$ [see Eq.~(\ref{eq-Chern-topological-charge})], which agrees with the Berry-curvature integral results reported in Ref.~\cite{10.1103/physrevlett.118.147401}.

The second example is the BHZ model. We consider the following Hamiltonian, $\mathcal{H}= \mathcal{H}_0 + \mathcal{H}_{V}$, with
\begin{equation}
\begin{aligned}
    \mathcal{H}_0
    &= \sum_{\mathbf{k}} \hat{c}^{\dagger}_{\mathbf{k}\mu} [H(\mathbf{k})]_{\mu\nu} \hat{c}^{\dagger}_{\mathbf{k}\nu},
    \\
    \mathcal{H}_{V}
    &= \frac{1}{2N^2}
    \sum_{\mathbf{q}\mathbf{k}\mathbf{k}^{\prime}}
    V_{\mathbf{q}}
    \hat{c}^{\dagger}_{\mathbf{k}-\mathbf{q},\mu}
    \hat{c}^{\dagger}_{\mathbf{k}^{\prime}+\mathbf{q},\nu}
    \hat{c}_{\mathbf{k}^{\prime}\nu}\hat{c}_{\mathbf{k}\mu}.
\end{aligned}
\end{equation}
Here $\mu$ involves two spinful $s$ orbitals ($|s\uparrow\rangle$, $|s\downarrow\rangle$) and two spin-orbit coupled $p$ orbitals ($|p_{+}\uparrow\rangle$, $|p_{-}\downarrow\rangle$) on a square lattice. $N$ is the number of unit cells.
The tight-binding Hamiltonian $H(\mathbf{k})=H_{\text{BHZ}}+H_{B_z}$ consists of the BHZ Hamiltonian $H_{\text{BHZ}}$, and a Zeeman term $H_{B_z}$ induced by a perpendicular magnetic field $(0,0,B_z)$, which incorporates both \textit{spin} and \textit{orbital} Zeeman couplings.
The Coulomb interaction potential is treated via a reciprocal-space Keldysh potential $V_{\mathbf{q}} = \frac{e^2}{2\epsilon_0 \bar{\epsilon}A} \frac{1}{q(1 + r_0 q)}$ \cite{PismaZhETF.29.716}. Details of the model and the numerical treatment are presented in Sec.~III.A of \cite{SI}.
Since $H_{B_z}$ preserves spin conservation, excitons can always be classified into three decoupled sectors according to their spin: $S_z=0,\pm1$, which allows us to study their dispersions and topological properties separately. Our investigation covers both the trivial and QSH insulating phases, with and without the Zeeman term $H_{B_z}$ (see Sec.~III.B of \cite{SI} for details).
This extends beyond the previous works by Blason and Fabrizio \cite{10.1103/physrevb.102.035146}, and Shindou, \textit{et al.} \cite{10.1103/physrevb.96.161101}. The former focused on the $S_z=\pm1$ exciton topology in the QSH phase of the BHZ model, while the latter studied the $S_z=1$ exciton topology in a two-band Chern insulator, which can be viewed as a magnetized variant of the full BHZ model.

We are particularly interested in the $S_z=0$ excitons in the QSH regime, as they provide an example of a multiple-BCP case. Fig.~\ref{fig-BHZ-exciton-dispersion} illustrates the $S_z=0$ exciton dispersions without and with $H_{B_z}$.
In the absence of $H_{B_z}$, the two lowest exciton bands, denoted by $|\Psi_{\mathbf{Q},\pm}\rangle$, touch at $\mathbf{Q}_0=\Gamma$ and $M$, giving rise to two BCPs [see Fig.~\ref{fig-BHZ-exciton-dispersion}(a)].
We note that the degeneracies of both BCPs are protected by fourfold rotational symmetry ($C_4$) and TRS (see Sec.~III.D of \cite{SI}).
The presence of TRS further enforces the topological triviality of $S_z=0$ excitons. While this has been previously discussed in Ref.~\cite{10.1103/physrevb.102.035146}, the case with broken TRS has not yet been explored.
Breaking TRS leads to two important changes. First, $|\Psi_{\mathbf{Q}_0,\pm}\rangle$ are no longer symmetry-protected to remain degenerate, thus allowing a gap to open at the BCPs. Second, the topology is no longer necessarily trivial. As shown in Fig.~\ref{fig-BHZ-exciton-dispersion}(b), both BCPs (marked by red stars) are simultaneously gapped out in the presence of $H_{B_z}$, yielding a fully isolated lowest exciton band. We note that, both degeneracy liftings are specifically driven by the \textit{orbital Zeeman coupling}, whereas the spin Zeeman coupling alone has no effect (see Sec.~III.C of \cite{SI}); this holds for all $S_z=0$ excitons in spin-conserved systems. Our numerical calculations show that this lowest exciton band acquires a nonzero exciton Chern number $\mathcal{C}_{\text{exc}}=2$ for $B_z>0$, indicating that breaking TRS drives the $S_z=0$ exciton into a topological phase.


\begin{figure}[tb!]
    \includegraphics[width=0.8\linewidth]{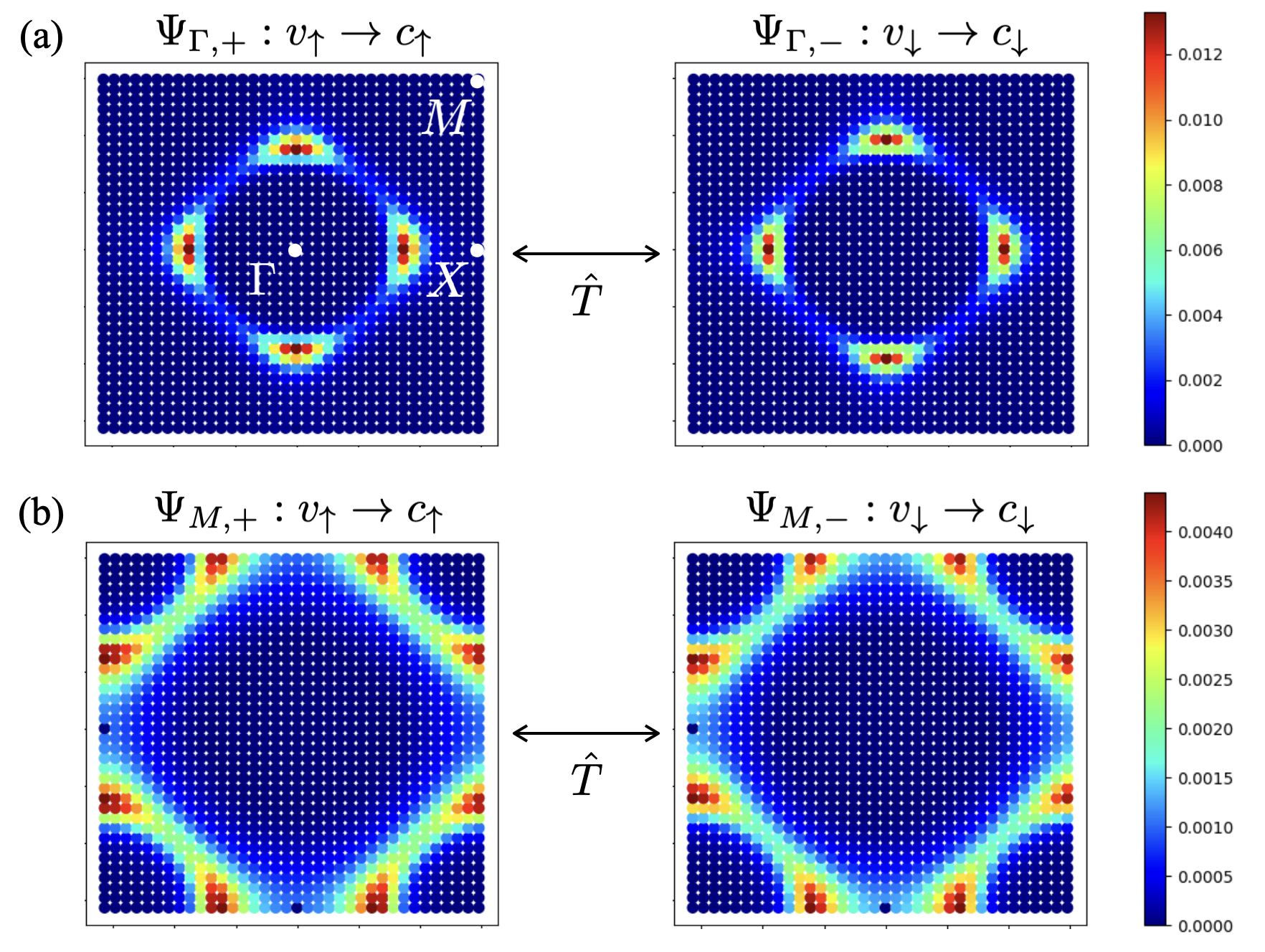}%
    \caption{Modulus squares of envelope functions $\phi_{\mathbf{Q}_0vc}(\mathbf{k})$ of $\Psi_{\mathbf{Q}_0,\sigma}$ at (a) $\mathbf{Q}_0=\Gamma$ and (b) $M$, in terms of the band-to-band transitions $v_{\sigma}\rightarrow{c}_{\sigma}$, where $\sigma=+(\uparrow),-(\downarrow)$ denotes two exciton states (the spin of electron band).
    All envelope functions respect fourfold rotational symmetry and are related by time-reversal symmetry ($\hat{T}$) in the absence of the Zeeman term \cite{figure}.
    We note that while $\Psi_{\Gamma,\sigma}$ only has contribution from $v_{\sigma}\rightarrow c_{\sigma}$, $\Psi_{M,\sigma}$ has minor contributions from $v_{\bar{\sigma}}\rightarrow c_{\bar{\sigma}}$, which can be neglected (see Sec.~III.D of \cite{SI}).
    }
    \label{fig-BHZ-exciton-b2b-con}
\end{figure}

\begin{figure}[tb!]
    \includegraphics[width=0.8\linewidth]{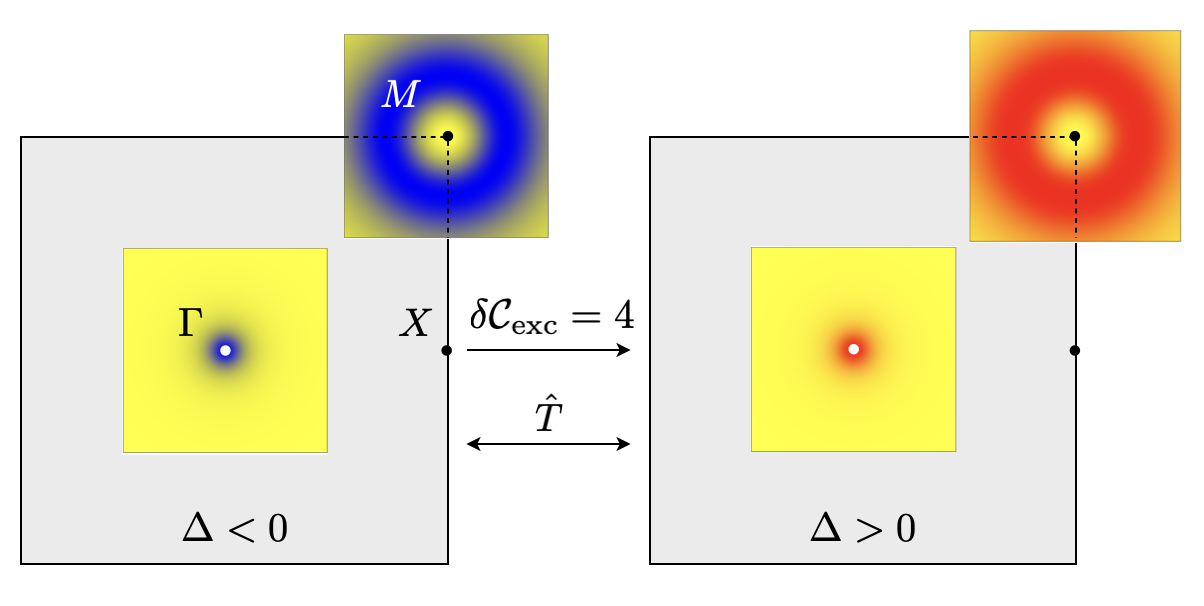}%
    \caption{Local Berry curvature distribution of the lowest $S_z=0$ exciton band in the BHZ model for (a) $\Delta<0$ and (b) $\Delta>0$.
    The exciton Chern number is $\mathcal{C}_{\text{exc}}(\Delta)=2\operatorname{sgn}(\Delta)$.
    }
    \label{fig-BHZ-Berry-curvature-pattern}
\end{figure}

The nontrivial topology of $S_z=0$ excitons induced by TRS-breaking stems from the local topological structures around two BCPs, which can be explicitly captured using our continuum theory approach.
The symmetry-protection mechanism of the both BCPs implies that the corresponding continuum Hamiltonians should respect the associated symmetries. In other words, they can be obtained via symmetry analysis and solved analytially.
Fig.~\ref{fig-BHZ-exciton-b2b-con} presents the modulus squares of the envelope functions of $|\Psi_{\mathbf{Q}_0,\pm}\rangle$ in terms of the band-to-band transitions, which clearly reflect the presence of both $C_4$ and TRS.
We also observe that, their dominant transitions are distributed along a square-shaped ring, rather than being localized around a specific point as in the exciton states studied in TMDs \cite{10.1103/physrevlett.108.196802, 10.1038/nphys3201, 10.1103/physrevb.91.075310, 10.1103/physrevlett.119.127403} and Bi/SiC \cite{10.1073/pnas.2307611120}. This feature originates from the band inversion mechanism of the single-particle system \cite{band-inversion}.
Constrained by $C_4$ and TRS, the continuum Hamiltonian near $\mathbf{Q}_0=\Gamma,M$ takes the form (details are present in Sec.~III.E of \cite{SI}):
\begin{equation}\label{eq-BHZ-exeffH}
\begin{aligned}
    &H^{\text{eff}}_{\mathbf{Q}_0}(\mathbf{Q})
    = \left[\hbar\Omega_0 + j(Q)\right]\cdot{I}_2
    \\
    &\quad\quad+
    j^{\prime}(Q)\left[
    \begin{array}{cc}
        1 & e^{-i2\phi_{\mathbf{Q}}}\\
        e^{i2\phi_{\mathbf{Q}}} & 1
    \end{array}
    \right]
    +
    \left[
    \begin{array}{cc}
        \Delta &\\
        & -\Delta
    \end{array}
    \right]
    ,
\end{aligned}
\end{equation}
where $j(Q)\propto{Q}^2$ and $j^{\prime}(Q)\propto{V}_{\mathbf{Q}_0+\mathbf{Q}}Q^2$.
The first two terms are the continuum Hamiltonian without the Zeeman contribution, while the last term accounts for the external Zeeman term (here $\Delta$ is tuned by $B_z$, with $\Delta=0$ when $B_z=0$).
Eq.~(\ref{eq-BHZ-exeffH}) indicates that both BCPs carry $N_3 = 2$.
In Fig.~\ref{fig-BHZ-Berry-curvature-pattern}, we also illustrate how the local topological structure around two $\mathbf{Q}_0$ points changes upon changing the sign of $\Delta$.
Consequently, the lowest exciton band has $\mathcal{C}_{\text{exc}}(\Delta)=2\operatorname{sgn}(\Delta)$, in agreement with our numerical results.
Finally, we note that the off-diagonal elements in Eq.~(\ref{eq-BHZ-exeffH}) arise solely from the exchange e-h interaction, which thus underlies the nontrivial topology. In contrast, the $S_z=\pm 1$ exciton topology originates differently from the direct e-h interaction (see Sec.~III.F of \cite{SI} for details).

\textit{Discussion and outlook.}---We developed a general continuum theory for exciton bands near $N$-fold BCPs in 2D systems. This framework applies broadly to BCPs located at both $\mathbf{Q}=0$ and finite $\mathbf{Q}$ points, and is valid for systems with or without spin conservation. Since the essential physics governing the exciton Chern number is localized near the BCPs, our theory provides a unified and physically transparent approach for characterizing exciton topology without requiring extensive numerical calculations.
In particular, we focused on systems with TRS and demonstrated that introducing a TRS-breaking parameter $\Delta$ can gap the BCPs and drive an initially topologically trivial system into a nontrivial phase. The resulting exciton Chern number is given by $\mathcal{C}_{\text{exc}}(\Delta) = \frac{1}{2} \sum_{i}w_{i} \cdot \operatorname{sgn}(\Delta)$, where $w_i$ is the winding number of the $i$th BCP.
Our result can be readily extended to a broad class of systems, such as twisted bilayer systems, Kane-Mele-type systems, and BHZ-type of quantum wells.

Our work ultimately traces back to our original attempt to investigate the excitonic properties in Bi/SiC. Our previous studies \cite{10.1126/science.aai8142,10.1103/physrevb.98.161407,10.1038/s41467-022-33822-8} revealed that, this system exhibits both global and local electron topology, as well as unique excitonic optical selection rules \cite{BiSiC}, all of which arise from the substrate-induced Rashba SOC.
A natural next step is to study the exciton dispersion and its associated topology.
While a full characterization of the global exciton topology remains challenging---due to the presence of multiple band crossings, a feature common in realistic materials \cite{10.1103/physrevb.91.075310,10.1103/physrevb.98.125206}---our continuum theory provides a reliable analytical framework to describe exciton behavior in the vicinity of specific $\mathbf{Q}$ points. This enables a clear characterization of the local exciton topology and, more broadly, offers a promising route toward unraveling the global topological structure of exciton bands in complex materials.

\textit{Acknowledgments.}—--A.C. acknowledges support from PNRR MUR project PE0000023-NQSTI.


\sloppy
\bibliography{reference}

\begin{thebibliography}{63}
\bibitem [{top()}]{topological-charge}%
  \BibitemOpen
  \href@noop {} {}\bibinfo {note} {Since the $N^{i}_3$ is a topological
  invariant, one can choose the closed surface $S_i$ to be a sufficiently small
  volume such that $\mathbf{\Omega}(Q_x,Q_y,\Delta)$ on the surface can be
  obtained by the effective continuum Hamiltonian near the BCP.}\BibitemShut
  {Stop}%
\bibitem [{che()}]{chern-diff}%
  \BibitemOpen
  \href@noop {} {}\bibinfo {note} {For instance, this method reveals that the
  (electron) Chern number of each spin branch in the BHZ model differs by
  $\pm1$ across the transition, indicating that one side is topologically
  nontrivial \cite{10.1126/science.1133734}. However, identifying which side is
  nontrivial still requires integration over the entire BZ.}\BibitemShut
  {Stop}%
\bibitem [{kin()}]{kink-state}%
  \BibitemOpen
  \href@noop {} {}\bibinfo {note} {This idea was originally proposed in the
  investigation of electron edge states in bilayer graphene
  \cite{10.1103/physrevlett.100.036804, 10.1103/physrevlett.102.096801,
  10.1073/pnas.1308853110}, and was later applied to the discussion of exciton
  topology in monolayer TMD \cite{10.1038/srep42390}. However, we note that
  Ref.~\cite{10.1038/srep42390} only considers the single BCP at the $\Gamma$
  point, whereas the exciton dispersion in monolayer TMDs exhibits multiple
  BCPs (see Ref.~\cite{10.1103/physrevb.91.075310}).}\BibitemShut {Stop}%
\bibitem [{iso()}]{isotropic-assumption}%
  \BibitemOpen
  \href@noop {} {}\bibinfo {note} {For instance, the exciton in TMD monolayers
  exhibits $w=2$ in pristine material \cite{10.1103/physrevb.91.075310} and
  $w=1$ under strain \cite{10.1038/ncomms4876}. Moreover, the $S_z=0$ exciton
  in the BHZ model has $w=2$, as we will discuss later [see
  Eq.~(\ref{eq-BHZ-exeffH})].}\BibitemShut {Stop}%
\bibitem [{fig()}]{figure}%
  \BibitemOpen
  \href@noop {} {}\bibinfo {note} {In the numerics, a small Zeeman term is
  included to lift the degeneracy between exciton states for clarity. As a
  result, the envelope functions are not strictly related by time-reversal
  symmetry, though their correspondence remains apparent.}\BibitemShut {Stop}%
\bibitem [{BiS()}]{BiSiC}%
  \BibitemOpen
  \href@noop {} {}\bibinfo {note} {While the global aspect is manifested in the
  emergence of the QSH insulating phase
  \cite{10.1126/science.aai8142,10.1103/physrevb.98.161407}, the local aspect
  is encoded in its valley physics, which further gives rise to an exciton
  optical selection rule \cite{10.1038/s41467-022-33822-8} --- and consequently
  a new type of exciton qubit \cite{10.1073/pnas.2307611120} --- that is
  fundamentally distinct from that of the widely studied TMDs.}\BibitemShut
  {Stop}%
\end{thebibliography}

\end{document}